\documentclass[letterpaper]{JHEP3}
\usepackage{epsfig}

\def\ba{\begin{eqnarray}}
\def\ea{\end{eqnarray}}
\def\be{\begin{equation}}
\def\ee{\end{equation}}
\def\nn{\nonumber}
\def\exd{{\rm d}}
\def\pd{\partial}
\makeatletter
\def\x@arrow{\DOTSB\Relbar}
\def\xlongequalsignfill@{\arrowfill@\x@arrow\Relbar\x@arrow}
\newcommand{\xlongequal}[2]{%
    \ext@arrow 0099\xlongequalsignfill@{#1}{#2}}
\makeatother

\newcommand{\roughly}[1]{\mathrel{\raise.3ex\hbox{$#1$\kern-0.85em
\lower1ex\hbox{$\sim$}}}}

\def\endignore{}
\def\ignore #1\endignore{} % use to "comment out" text

\def\be{\begin{equation}}
\def\beq\begin{equation}
\def\ee{\end{equation}}
\def\bea{\begin{eqnarray}}
\def\eea{\end{eqnarray}}

\def\nn{\nonumber}

\def\pref#1{(\ref{#1})}

\def\beq{\begin{equation}}
\def\eeq{\end{equation}}
\def\beqa{\begin{eqnarray}}
\def\eeqa{\end{eqnarray}}

\def\cA{{\cal A}}
\def\cB{{\cal B}}

\def\cF{{\cal F}}

\def\cL{{\cal L}}

\def\cR{{\cal R}}

\def\ssA{{\scriptscriptstyle A}}
\def\ssB{{\scriptscriptstyle B}}
\def\ssC{{\scriptscriptstyle C}}
\def\ssD{{\scriptscriptstyle D}}
\def\ssE{{\scriptscriptstyle E}}
\def\ssG{{\scriptscriptstyle G}}
\def\ssH{{\scriptscriptstyle H}}

\def\ssM{{\scriptscriptstyle M}}
\def\ssN{{\scriptscriptstyle N}}
\def\ssP{{\scriptscriptstyle P}}
\def\ssQ{{\scriptscriptstyle Q}}
\def\ssR{{\scriptscriptstyle R}}
\def\ssS{{\scriptscriptstyle S}}
\def\ssT{{\scriptscriptstyle T}}
\def\ssU{{\scriptscriptstyle U}}
\def\ssV{{\scriptscriptstyle V}}
\def\ssW{{\scriptscriptstyle W}}

\newcommand{\bmat}{\left(\begin{array}}
\newcommand{\emat}{\end{array}\right)}

\def\-{\hphantom{-}}

\def\s2{\frac{1}{2}}

\def\IF{\relax{\rm I\kern-.18em F}}
\def\II{\relax{\rm I\kern-.18em I}}
\def\IP{\relax{\rm I\kern-.18em P}}
\def\IC{\relax{\rm I\kern-.48em C}}
\def\IR{\relax{\rm I\kern-.18em R}}
\def\IK{\relax{\rm I\kern-.20em K}}
\def\IM{\relax{\rm I\kern-.25em M}}

\def\Dsl{\,\raise.15ex\hbox{/}\mkern-13.5mu D}
\def \one{\relax{\rm 1\kern-.26em I}}

\def\exd{{\rm d}}

\def\V{\mathcal{V}}

\def\nn{\nonumber}

\def\({\left(}
\def\){\right)}

\title{On  Brane Back-Reaction and de Sitter Solutions in Higher-Dimensional Supergravity}

\author{ C.P.~Burgess,${}^{1,2}$ Anshuman Maharana,${}^3$ L. van Nierop,${}^1$
A. A. Nizami${}^3$ and F.~Quevedo${}^{3,4}$\\

${}^1$Department of Physics \& Astronomy, McMaster University\\ \qquad 1280 Main Street West, Hamilton ON, Canada.%\\

${}^2$Perimeter Institute for Theoretical Physics\\
\qquad 31 Caroline Street North, Waterloo ON, Canada.%\\

$^3$ DAMTP/CMS, University of Cambridge, %Wilberforce Road,\\
 Cambridge CB3 0WA, UK.%\\

$^4$ Abdus Salam ICTP, Strada Costiera 11, Trieste 34014, Italy.
}

\date{}
\abstract { We argue that the problem of finding lower-dimensional de Sitter solutions to the classical field equations of higher-dimensional supergravity necessarily requires understanding the back-reaction of whatever localized objects source the bulk fields. However, we also find that most of the details of the back-reacted solutions are not important for determining the lower-dimensional curvature. We find, in particular, a classically exact expression that, for a broad class of geometries, directly relates the curvature of the lower-dimensional geometry to asymptotic properties of various bulk fields near the sources. Specializing to codimension-two sources, we find that the contribution involving the asymptotic behaviour of the warp factor (which has a definite sign for most supergravities and so is usually used to infer a preference for anti-de Sitter geometries) is precisely canceled by the contribution of the sources themselves (that are left out in earlier treatments). We identify which combination of bulk fields survives this cancelation, and so controls the sign of the lower-dimensional geometry, for several supergravities in 6, 10 and 11 dimensions. Our results show precisely why explicit 4D de Sitter solutions to 6D supergravity evade general no-go theorems. As an application we show that {\em all} classical compactifications of Type IIB supergravity  (and F-theory) to 8 dimensions are 8D-flat if they involve only the metric and the axio-dilaton sourced by codimension-two sources, extending earlier results to include warped solutions and more general source properties.
}

\preprint{DAMTP-2011-66}

\begin{document}
\section{Introduction}

de Sitter space, or slow-roll geometries close to de Sitter space, appear to play an important role in cosmology. For those who believe that extra dimensions exist this has motivated searching for explicit solutions to the higher-dimensional field equations for which the large four dimensions we see are de Sitter or de Sitter-like. Although a few such solutions are known \cite{CG, 6DdS}, more and more general no-go results \cite{GG, MN, Wox, StW} show that such solutions are difficult to find\footnote{Four-dimensional effective field theories of string theory including non-perturbative effects and anti branes or D-terms \cite{eft}\  can give rise to de Sitter solutions. But at the moment there is no full understanding from the  microscopic higher-dimensional theory. For other recent attempts for de Sitter solutions see \cite{attempts}.} Why should this be so?

In this paper we argue that part of the problem is that we are not yet using all of the ingredients that de Sitter solutions require. In particular, contributions are being neglected that are the same size as some of the contributions that are usually kept when searching for (or ruling out) de Sitter-like solutions.

The neglected contributions come from the actions of any localized sources that may be present in the extra-dimensional configurations of interest. In particular, we argue here that for codimension-two sources these actions contribute to the curvature an amount that is competitive with the contribution of the bulk fields, including their back-reaction. In particular, the source action acts to systematically cancel the contribution from the warping of the noncompact geometry across the extra dimensions. This is important because the sign of the warping contribution is usually definite, and because it is opposite to what is required for a de Sitter noncompact geometry it plays a role in the various extant de Sitter no-go results.

\subsection{No-go results and the 6D loophole}

Our interest is in $D$-dimensional metrics of the form
\be \label{metricform}
 \exd s^2 = \hat g_{\ssM \ssN} \, \exd x^\ssM \exd x^\ssN
 = e^{2W(y)} \,  g_{\mu\nu}(x) \, \exd x^\mu \exd x^\nu + \tilde g_{mn}(y) \, \exd y^m \exd y^n \,,
\ee
where $D = d + n$; the $d$-dimensional metric, $g_{\mu\nu}$, is maximally symmetric ({\em i.e.} flat, de Sitter or anti-de Sitter); and the warp factor, $W$, can depend on position in the $n$ compact directions (whose metric, $\tilde g_{mn}$, is so far arbitrary).

In particular, for cosmological applications there is much interest in identifying solutions to higher-dimensional field equations for which $g_{\mu\nu}$ is a de Sitter metric (which in our curvature conventions\footnote{We use a `mostly plus' metric and Weinberg's curvature conventions \cite{Wbg}, which differ from those of MTW \cite{MTW} only in the overall sign of the definition of the Riemann tensor.} satisfies $R = g^{\mu\nu} R_{\mu\nu} < 0$). The search for such solutions has been fairly barren, and this is partly explained by refs.~\cite{GG}, \cite{MN}, \cite{Wox} and \cite{StW}, who identify increasingly general obstacles to finding this type of de Sitter solution to sensible, higher-dimensional, second-derivative field equations.

On the other hand, a handful of explicit solutions of this type {\em do} exist, including 4D de Sitter solutions \cite{CG} for six-dimensional Maxwell-Einstein systems,
\be \label{MEAction}
 S_{\ssM\ssE} = - \int \exd^6 x \sqrt{- \hat g} \; \left\{ \frac1{2\kappa^2} \, \hat g^{\ssM \ssN} \hat \cR_{\ssM \ssN}
 + \frac14 \, \cF_{\ssM\ssN} \cF^{\ssM\ssN} + \Lambda
 \right\} \,,
\ee
with positive 6D cosmological constant, $\Lambda$. Similar solutions \cite{6DdS} also exist for six-dimensional gauged, chiral supergravity \cite{NS}, whose relevant bosonic action is
\be \label{NSAction}
 S_\mathrm{bulk} = - \int \exd^6 x \sqrt{- \hat g} \; \left\{ \frac1{2\kappa^2} \, \hat g^{\ssM\ssN}
 \Bigl( \hat \cR_{\ssM \ssN}
 + \pd_\ssM \phi \, \pd_\ssN \phi \Bigr)
 + \frac14 \, e^{-\phi} \cF_{\ssM\ssN} \cF^{\ssM\ssN}
 + \frac{2 \, g_\ssR^2}{\kappa^4} \, e^\phi \right\} \,.
\ee
For both of these actions $\hat \cR_{\ssM\ssN}$ denotes the Ricci tensor for the 6D metric, $\hat g_{\ssM \ssN}$, and $\cF = \exd \cA$ is the field strength for a 6D gauge potential, $\cA_\ssM$. The quantity $\kappa^2 = 8\pi G_6$ denotes the 6D gravitational coupling, while for the supersymmetric case $g_\ssR$ denotes the gauge coupling of a specific $U_\ssR(1)$ gauge group that does not commute with 6D supersymmetry.

These examples do not contradict the various no-go theorems because they arise in systems which do not satisfy one of the assumptions of each. For instance, the no-go result of \cite{MN} assumes that any extra-dimensional scalar potential must be negative (as it tends to be for higher-dimensional supergravities, but is not so for eqs.~\pref{MEAction} and \pref{NSAction}). They evade the less restrictive assumptions of \cite{Wox} and \cite{StW}, some of which exclude \cite{StW} having only two extra dimensions, $n = 2$. More importantly, for this paper, they do not satisfy the average `boundedness' assumptions \cite{Wox} that exclude solutions that are too singular.

\subsection{The potential relevance of back-reaction}

There are two ways to view the possibility that singular behaviour can suffice to evade the no-go results. One view is to regard solutions with such singularities as unacceptable, and so draws the conclusion that de Sitter solutions may be impossible to find. And for some types of singularity (like negative-mass black holes) this is probably right, since the alternative requires admitting energies that are unbounded from below.

But some (apparent) singularities are known to be perfectly sensible, such as those seen in Coulomb's law at the position of a source charge. In the case of Coulomb's law, the singularity doesn't preclude taking the solution seriously because we don't intend to trust the solution in any case right down to zero size. The existence of apparent singularities might similarly be expected to arise in the gravitational theories relevant to cosmology, provided these are regarded as effective descriptions of some more-microscopic degrees of freedom. One can hope to get a handle on deciding whether a singularity might be reasonable for an effective description, by seeing what kinds of apparent singularities actually can emerge from localized sources governed by physically reasonable actions.

These considerations suggest that understanding the back-reaction of localized sources could be a crucial part of obtaining de Sitter solutions, or ruling them out. In particular the asymptotics, and apparent divergence, of bulk fields near a source is likely to be important, and is ultimately controlled by the action that describes the dynamics of that source. Notice for these purposes `source' need not mean a fundamental object, like a D-brane. Rather, it could describe something more complicated, like a soliton or a higher-dimensional brane wrapping internal dimensions, a localized but strongly warped region, or a more complicated object (like a nucleus or a star). All we need know is that the sources are much smaller than the extra dimensions within which they sit.

How the properties of a source affect the properties of bulk fields is best understood at present for codimension-one and codimension-two sources. For codimension-one sources, the back-reaction is described by the Israel junction conditions \cite{IJC}, as is familiar from Randall-Sundrum models \cite{RS}. But bulk fields with codimension-one sources also tend not to diverge at the source positions, and so shed little light on how such singularities influence the low-energy curvature. It is only for higher-codimension sources that it is generic that bulk fields diverge at the source positions, and so where the relation between bulk singularity and source properties can be explored.

Of course, these bulk singularities make matching bulk solutions to source properties more complicated, usually requiring a renormalization of the source \cite{Bren}. The tools for detailed bulk-source matching and renormalization are most explicitly known for codimension-two objects \cite{Cod2Matching, otheruvcaps, BBvN, BulkAxions, susybranes}. In particular, these tools have recently been used to identify \cite{6DInf} explicit objects that can source the de Sitter solutions \cite{6DdS} of the 6D supergravity action, eq.~\pref{NSAction}. Since the required source properties seem physically reasonable,\footnote{As discussed in more detail below, their worst feature appears to be a requirement that the dilaton, $\phi$, grows as one asymptotically approaches the sources, and so care must be taken to avoid leaving the weak-coupling regime before reaching the source.} they show that the singularities in the corresponding bulk solutions need not be regarded as grounds for their rejection.

\subsection{Summary of results}

In the rest of this paper we examine how source back-reaction constrains the existence of de Sitter solutions in more general higher-dimensional theories than the six-dimensional ones already explored.

In particular, we explore some of these issues in eleven-dimensional supergravity, and in ten-dimensional Type IIB and Type IIA supergravity. Because our best-developed tools apply to codimension-two objects, it is these we largely explore in detail. If only $D$-branes were allowed as sources, this would restrict us to $D7$-branes in Type IIB systems. But we also explore the other supergravities for two reasons: because some of our results apply equally well to higher-codimension sources; and because our sources might not be $D$-branes --- or $(p,q)$ branes for that matter --- but instead be more complicated localized codimension-two quantities (like very small warped throats).

We find the following results:
\begin{itemize}
\item First, for geometries of the form of eq.~\pref{metricform}, we find a very general classical relationship that gives the curvature in the non-compact dimensions parallel to the sources as the sum of four terms: $R \propto I + II + III + IV$, where $IV$ vanishes for maximally symmetric geometries in the absence of space-filling fluxes.
\item Second, we show that contribution $I$ --- which is proportional to the bulk action evaluated at the classical back-reacted solution --- is very generally given as the integral of a total derivative, and so is controlled by the boundary values of a particular combination of bulk fields. This property relies only on the existence of a classical scale invariance that is shared by most higher-dimensional supergravities (and holds in particular for 11D and 10D Type IIA and IIB supergravity).
\item Third, we show that for codimension-two sources the contributions $II$ and $III$ cancel one another. Here contribution $II$ is an integral over a total derivative of the warp factor, $W$, whose definite sign plays an important role in the derivation of the general no-go results. Contribution $III$ comes from the action of the localized source, which is left out of most no-go analyses.
\item Finally, we explicitly identify the total derivative that appears in $I$ for several examples of interest, including commonly used supergravities in 6, 10 and 11 dimensions. This identifies the combination of fields whose near-brane asymptotics is relevant to the low-energy curvature. As a simple application we show that the noncompact dimensions are always flat for all F-theory compactifications that involve only the metric and axio-dilaton with codimension-two sources.
\end{itemize}

These results carry two important messages. First, since the direct contributions from the source action cancel important contributions in the no-go theorems, the bad news is that back-reaction cannot be neglected when determining the curvature of the noncompact dimensions. But second, because the nonzero contributions are total derivatives, the good news is that most of the details of the back-reacted solutions are {\em not} important. All that counts is the near-source asymptotics of a specific combination of back-reacted bulk fields.

Our explanation of these results is organized as follows. The next section, \S2, develops general expressions for how the curvature of non-compact, maximally symmetric directions depends on the properties of the extra-dimensional bulk fields. Much of this section is similar in spirit to the arguments made when deriving no-go results \cite{GG, MN, Wox, StW}, and our main new contribution is to cleanly identify how the curvature is controlled by asymptotic forms near the sources, and to see how assumptions about source dynamics modifies this asymptotics. \S2\ also explicitly identifies for 11D and 10D supergravity the precise combination of bulk fields whose asymptotic forms are relevant to the low-energy curvature. \S3\ then applies these general arguments to the special case of metric/axio-dilaton configurations in 10D Type IIB supergravity with codimension-two sources, showing in this case how all solutions are flat in the noncompact directions in the absence of bulk fluxes. We summarize our conclusions in \S4, and several appendices provide details of calculations used in the main text.

\section{Low energy curvature and near-source asymptotics}

The purpose of this section is to derive a general expression for the curvature of the noncompact directions that is our main result. We do so by paralleling arguments made elsewhere for six-dimensional supergravities \cite{6DdS, Cod2Matching, BBvN}.

We make the connection between on-source curvatures and near-source asymptotics in three steps. First, in \S\ref{seco} we show --- at the classical level for maximally symmetric source geometries --- that the integral of the low-energy curvature can be computed as the sum of four terms: $I + II + III + IV$. Of these, $I$ is the higher-dimensional bulk action, evaluated at the compactified solution. $II$ is the integral over a total derivative, which Gauss' theorem directly relates to the boundary values of the warp factor, at infinity and near any potential singularities. $III$ is a direct contribution from the action of any sources, and $IV$ is a term which vanishes in the absence of any space-filling fluxes.

Next, the second step is taken in \S\ref{sect}, which shows that for all of the supergravities of interest the higher-dimensional bulk lagrangian density is itself also always a total derivative when evaluated at an arbitrary classical solution. Combining this with step one then shows that, in the absence of space-filling fluxes, the integrated low-energy curvature is completely controlled by source and boundary effects.

Finally, \S\ref{secpb} demonstrates step three. By treating carefully the singular behaviour near any codimension-two sources, it is shown that contributions $II$ and $III$ precisely cancel one another. Taken together, these three steps show that only contribution $I$ plays any role in a broad class of theories.

\subsection{Step 1: Integrating out the bulk}
\label{seco}

We first focus on step one: we use the higher dimensional equations of motion to derive a relationship between the lower dimensional curvature and the on-shell higher-dimensional action. For definiteness, we consider solutions to the field equations of a $D$-dimensional (super)gravity theory, with action\footnote{An aside on notation: indices $M,N = 0,1, \dots, D-1$ run over all dimenesion; greek indices denote lower-dimensional coordinates $\mu, \nu = 0,1, \dots, d-1$; and indices $m,n = 1, \dots, n = D-d$ denote compactified coordinates. We use $\hat \cR_{\ssM \ssN}$ to denote the $D$-dimensional Ricci curvature of the full $D$-dimensional metric, $\hat g_{\ssM \ssN}$; and $\hat R_{\mu\nu}$ to denote the $d$-dimensional Ricci curvature computed from the $d$-dimensional metric, $\hat g_{\mu\nu} = e^{2W} g_{\mu\nu}$. Finally, $\hat g_\ssD = \det \hat g_{\ssM \ssN}$ while $\hat g_d = \det \hat g_{\mu\nu}$ {\em etc.}}
\be \label{GenDAction}
   S = { 1 \over 2 \kappa_{\ssD}^{2} }
   \int \exd^{\ssD} z  \sqrt{-\hat{g}_{\ssD}}
   \bigg( - \hat{{\cal{R}}}
   +{\cal{L}}_{\rm {matter}}^{\ssD} \bigg)
   + S_{\rm source} \,,
\ee
where $\cL_{\rm matter}$ depends on a generic set of other $D$-dimensional fields (but not on the derivatives of the metric), denoted collectively by $\psi$. $S_{\rm source}$ denotes the action of any sources, which differs from the term explicitly written by only involving an integration over $d$ dimensions, rather than $D$.

Now imagine we have a solution to the field equations for this action describing a compactification down to $0 < d = D -n$ dimensions, of the form of eq.~\pref{metricform}. We wish to derive a general expression for $R = g^{\mu\nu} R_{\mu\nu}$ in terms of properties of the warp-factor, $W$, the compact metric, $\tilde g_{mn}$, and the bulk- and source-matter actions.

To this end consider the $\mu \nu$ component of Einstein's equation,
\begin{equation}
   \sqrt{-\hat{g}_{\ssD}}
   \left[ \hat{\cR}^{\mu \nu} + {1 \over 2 } \, \hat{g}^{\mu \nu}  \Bigl( -\hat{\cR} + {\cal{L}}^{\ssD}_{\rm matter} \Bigr) + { \partial {\cal{L}}_{\rm matter}^{\ssD} \over \partial \hat{g}_{\mu \nu} } \right]
   + 2 \kappa_\ssD^2 \, \left( \frac{\delta S_{\rm source}}
   {\delta \hat g_{\mu\nu}} \right) = 0 \,,
\end{equation}
which we contract with $\hat{g}^{\mu \nu}$, making use of
\ba
 \hat{g}^{\mu \nu}  \hat{\cR}_{\mu \nu}
 &=& e^{-2W} {{R}} + d \, \tilde{\nabla}^{2} W
 + d^2 \, \tilde g^{mn}
 \partial_m W \partial_n W  \nn\\
 &=& e^{-2W} {{R}} + e^{-d W} \tilde
 \nabla^2 e^{dW} \,,
\ea
where $\tilde \nabla^2 = \tilde g^{mn} \tilde\nabla_m \tilde\nabla_n$. Dividing the result by $ 2 \kappa_{\ssD}^{2}$, using $\sqrt{-\hat g_\ssD} = e^{dW} \sqrt{-g_d} \; \sqrt{\tilde g_n}$, and integrating over all $D$ dimensions then gives
\ba
\label{higher}
 -{ 1 \over 2 \kappa_{d}^{2} }
 \int \exd^{d} x \sqrt{-{g}_{d}} \; R
 &=& {d \over 2 } \; S_{\rm on-shell}
  + { 1 \over 2 \kappa_{\ssD}^{2} }
  \int \exd^{d} x \sqrt{-{g}_{d}} \; \int \exd^{n} y \sqrt{\tilde{g}_{n}} \; \tilde{\nabla}^2  e^{dW} \\
 && \quad  + \int \exd^d x \; \hat g_{\mu\nu}
  \left( \frac{\delta S_{\rm source}}
 {\delta \hat g_{\mu\nu}} \right)
 + { 1 \over 2 \kappa_{\ssD}^{2} }
 \int \exd^{\ssD}x  \sqrt{-\hat{g}_{\ssD}} \;
 \hat{g}_{\mu \nu}{ \partial
 {\cal{L} }^{\ssD}_{\rm matter} \over
 \partial \hat{g}_{\mu \nu} } \nn\\
 &:=& I + II + III + IV \,, \nn
\ea
where $S_{\rm on-shell}$ means the bulk part of the action appearing in eq.~\pref{GenDAction}, evaluated at a solution to the field equations, and the last term uses that the source terms are localized within the extra dimensions. $\kappa_d^2$ denotes the $d$-dimensional gravitational coupling given by $\kappa_d^2 = \kappa_\ssD^2/\V_\ssW$, with the warped volume defined by
\be \label{VWdef}
   \V_{\ssW} := \int \exd^{n} y \sqrt{\tilde{g}_n} \; e^{(d-2)W} \,.
\ee

\subsubsection*{Maximal symmetry and space-filling fluxes}

Eq.~\pref{higher} is the key equation, and so far it has been derived on very general grounds. We now specialize to the situation where the solution does not break the maximal symmetry of the $d$-dimensional metric $g_{\mu\nu}$.

Maximal symmetry is a very constraining condition. First, it implies $R$ is a constant, so the left-hand-side of eq.~\pref{higher} is proportional to the (divergent) volume of the noncompact dimensions. Furthermore, the left-hand-side vanishes only for flat $d$-dimensional space, and its sign is controlled by the sign of $R$.

Second, maximal symmetry strongly restricts the form of $\partial \cL^\ssD_{\rm matter}/\partial \hat g_{\mu\nu}$ for the field content usually found in higher-dimensional supergravity. In particular, the only fields that can be nonzero (classically) for maximally symmetric solutions are: the metric, $g_{\mu\nu}$; space-filling fluxes of the form
\be
    F^{(p)}_{\mu_1 .. \mu_d m_1 .. m_{p-d}}
    = \epsilon_{\mu_1 .. \mu_d} G_{m_1 ... m_{p-d}} \,;
\ee
and any number of $d$-dimensional scalar fields (like components of $\tilde g_{mn}$, {\em etc.}).

Because $\cL^\ssD$ is defined with an overall factor of $\sqrt{- \hat g_\ssD}$ factored out, and because the Einstein term is also treated separately, in the absence of higher-derivative interactions $\partial \cL^\ssD_{\rm matter}/\partial \hat g_{\mu\nu} = 0$ if only scalar fields and the metric are present. For the supergravities of interest here the only nonvanishing contributions to ${ \partial {\cal{L} }^{\ssD}_{\rm matter} / \partial \hat{g}_{\mu \nu} }$ arise from $p$-form fields (with $p \geq d$), having nonzero space filling components.

For instance, for a $p$-form field with kinetic term
\begin{equation}
   {\cal{L} }^{\ssD}_{\rm p-form} = - {1 \over 2\, p ! } \, F_{(p)}^{2} \,,
\end{equation}
and non-vanishing space filling components we have
\be \label{spacef0}
 \hat g_{\mu\nu}
  { \partial {\cal{L} }^{\ssD}_{\rm matter} \over \partial \hat{g}_{\mu \nu} } =
   -{d \over 2(p-d)! } \;
  G_{ m_1 ..m_{p-d}} G_{ n_1 ..n_{p-d}}
 \tilde{g}^{{m}_{1}{n}_{1} } \tilde{g}^{{m}_{2}{n}_{2}}
 \cdots \hat{g}^{{m}_{p-d}{n}_{p-d} } = -{ d \, G^2 \over 2
 (p-d)! }  \,,
\ee
which contributes to the right-hand-side of eq.~\pref{higher} the amount
\be
\label{spacef}
 - { d \over 2 \kappa_{\ssD}^{2} (p-d)!} \int \exd^{d} x  \sqrt{-{g}_{d}} \int d^{n} y \sqrt{{\tilde{g}}_{n}}
 \; e^{dW} \, G^2 \,.
\ee
We note that this is negative definite, which (in our conventions) contributes to $R$ with an anti-de Sitter-like sign.

Of course, space-filling fluxes need not contribute to eq.~\pref{higher} only through their kinetic term. The quantity ${ \partial {\cal{L} }^{\ssD}_{\rm matter} / \partial \hat{g}_{\mu \nu} }$ can also receive contributions from Chern-Simons terms. In this case, because $\cL_{\rm CS \, matter}^\ssD = \cL_{\rm CS}/\sqrt{- g_\ssD}$, the contribution is simply proportional to the Chern-Simons term itself:
\begin{equation}
   \hat g_{\mu\nu}
  { \partial {\cal{L} }^{\ssD}_{\rm CS \, matter} \over \partial \hat{g}_{\mu \nu} } = - {d \over 2} \int \cal{L}_{\rm CS} \,.
\end{equation}
Unlike for the kinetic term, this contribution can have indefinite sign.

We see that in the absence of space-filling flux, the last term in equation (\ref{higher}) vanishes. When this is so, eq.~(\ref{higher}) relates the $d$-dimensional curvature, $R$, to a total derivative, a derivative of the source action, and the bulk action evaluated on shell (which we show below is often also a total derivative).

The restriction to no space-filling fluxes is also not very restrictive, because one can usually (Hodge) dualize a flux to get rid of any space filling components. But there can be some situations where this cannot be done, such as when the flux in question is the self-dual five form of Type IIB supergravity. In this case the self-duality condition relates the flux components in the internal and space-time directions. Appendix \ref{exam} uses several well-known examples to illustrate how eq.~\pref{higher} works in practice (in the absence of source terms), with and without space-filling flux.

\subsection{Step 2: A general expression for $S_{\rm on-shell}$}
\label{sect}

This section now proves that $S_{\rm on-shell}$ can generally also be expressed as the integral of a total derivative for the bulk supergravities of general interest.

This is actually a special case of a more general result \cite{WBW6D} that states that any scale-invariant system has this property, as we review here. It is generic to higher-dimensional supergravities because these typically all have a classical scale invariance \cite{WBFme}.

Consider therefore a generic collection of fields, $\varphi^i$, described by a lagrangian density that scales as $\cL \to s^p \, \cL$ when the fields scale as $\varphi^i \to s^{a_i} \, \varphi^i$, for some constants $s$, $p$ and $a_i$:
\be \label{scalingdef}
 \cL\left( s^{a_i} \varphi^i, s^{a_i} \partial_\mu \varphi^i \right) \equiv s^p \, \cL \left( \varphi^i , \partial_\mu \varphi^i \right) \,.
\ee
This scaling property of the action ensures the invariance of the field equations.

Eq.~\pref{scalingdef} should be read as being an identity for all $s$ and for all fields $\varphi^i$. Differentiating with respect to $s$ and evaluating the result at $s = 1$ then gives the identity
\be \label{Lvarn}
 \sum_i a_i \left[ \left( \frac{\partial \cL}{\partial \left[ \partial_\mu \varphi^i
 \right] } \right) \partial_\mu \varphi^i
 + \left( \frac{\partial \cL}{ \partial \varphi^i}
 \right) \varphi^i \right] = p \, \cL \,,
\ee
for all $\varphi^i$. But solutions to the field equations satisfy
\be
 \left( \frac{\partial \cL}{\partial \varphi^i } \right)
 - \partial_\mu \left( \frac{\partial \cL}{\partial \left[
 \partial_\mu \varphi^i \right] } \right) = 0 \,,
\ee
and so using this in eq.~\pref{Lvarn} implies
\be \label{Lvarn2}
  \cL_{\rm on-shell}
 = \sum_i \frac{a_i}{p} \,\partial_\mu \left[ \left( \frac{\partial \cL}{\partial
 \left[ \partial_\mu \varphi^i \right] } \right)  \varphi^i \right] \,.
\ee
That is, the lagrangian evaluates to a total derivative at {\em any} classical solution.

We next pause to record the explicit form for the total derivative for the 6D supergravity for which de Sitter solutions are known to exist, and for the 11D and 10D supergravities of more general interest here. The details of these evaluations are given in Appendix \ref{app:onshell}.

\subsection*{6D supergravity}

As a point of reference, we restate here the on-shell action as computed \cite{WBW6D} for chiral, gauged supergravity \cite{NS} in six dimensions. The relevant bosonic action, $S^6$, is given in eq.~\pref{NSAction} and scales as $S^{6} \rightarrow s^{2} \, S^{6}$ when $\hat g_{\ssM \ssN} \to s \hat g_{\ssM \ssN}$ and $e^{-\phi} \to s \, e^{-\phi}$. The on-shell lagrangian is therefore a total derivative, and is seen by explicit evaluation to be
\be
 S^{6}_{\rm on-shell} = \frac{1}{2 \kappa_{6}^{2}}
 \int \exd^6 x \; \sqrt{- \hat g_6} \; \Box \phi \,.
\ee

In our conventions, when used in eq.~\pref{higher}, this shows that an AdS sign corresponds to $\phi$ decreasing near the source, while a de Sitter sign arises when $\phi$ increases towards the source (a property that may also be directly verified of the explicit de Sitter solutions \cite{6DdS, 6DInf}). Since $e^{2\phi}$ counts loops in this system, consistency of the classical approximation requires that one encounters the physics that regulates the source before leaving the weak-coupling regime $e^{\phi} \ll 1$. Although this sounds worrisome, similar considerations apply to the gravitational field of a macroscopic source like the Earth. The large curvatures encountered if this field were extrapolated to zero size would also eventually invalidate a semiclassical approximation; but are not a problem in practice due to the prior intervention of the Earth's surface.

\subsection*{11D supergravity}

For 11D supergravity the bosonic action is
\be
 S^{11} = -\frac{1}{2 \kappa_{11}^{2}}
 \int \exd^{11} x \sqrt{- \hat g_{11}} \; \left[
 \hat \cR + \frac{1}{2(4!)} \, G_{4}^{2} \right]
 - {1 \over 12 \kappa_{11}^{2}}
 \int G_{4} \wedge G_{4} \wedge C_{3} \,.
\ee
This scales as $S^{11} \rightarrow s^{9/2} S^{11}$ when $\hat g_{\ssM \ssN} \to s \hat g_{\ssM \ssN}$ and $C_{\ssM \ssN \ssP} \to s^{3/2} C_{\ssM \ssN \ssP}$.

As argued above, this scaling behaviour implies that the on-shell lagrangian is a total derivative. Using the field equations gives the following total derivative expression for the on-shell 11D action:
\be
 S^{11}_{\rm on-shell} = -\frac{1}{6 \kappa_{11}^{2}}
 \int \exd \Bigl(C_{3} \wedge *G_{4} \Bigr) \,.
\ee
We note that the expression has explicit dependence on the potential $C_3$, thus
one can get non-trivial contributions from the patching of gauge charts. We
hope to explore such contributions in the future.
\subsection*{10D Type IIA supergravity}

The story for the 10D Type IIA supergravity action is similar. The Einstein-frame action for the bosonic sector is
\be
 S^{\rm IIA} = - \frac{1}{2 \kappa_{10}^{2}}
 \int \exd^{10} x \sqrt{-\hat g_{10}} \;
 \left[ \hat \cR + \frac{1}{2} \, (\partial\phi)^{2}
 + \frac{e^{-\phi}}{2(3!)} \, H_{3}^{2}
 + \frac{e^{3\phi/2}}{2(2!)} \, F_{2}^{2}
 + \frac{e^{\phi/2}}{2(4!)} \, \tilde{F}_{4}^{2} \right] +S_{CS} \,
\ee
with Chern-Simons term given by
\be
 S_{CS} = - {1 \over 4 \kappa_{10}^{2}} \int \exd C \wedge
 \exd C \wedge B  \,.
\ee

This action scales as $S^{\rm IIA} \to s^2 \, S^{\rm IIA}$ under the transformations $e^{-\phi} \to s \, e^{-\phi}$, $\hat g_{\ssM \ssN } \to \sqrt{s} \; \hat g_{\ssM \ssN}$, $C_k \to s \, C_k$ and $B_2 \to B_2$. This ensures the action can be written as a total derivative using the form-field equations of motion; explicitly
\ba
 S^{\rm IIA}_{\rm on-shell} &=& -\frac{1}{8 \kappa_{10}^{2}}
 \int \exd \left( - e^{-\phi} \, B_{2} \wedge *H_{3}
 + \frac{e^{3\phi/2}}{2} \, C_{1} \wedge *F_{2}
 + \frac{3e^{\phi/2}}{2} \, C_{3} \wedge *\tilde{F}_{4}
 \right. \\
 && \qquad\qquad\qquad\qquad\qquad\qquad
  \left. - e^{\phi/2} \, B_{2}
 \wedge C_{1} \wedge *\tilde{F}_{4}
 + \frac{3}{2} \, C_{3} \wedge F_{4} \wedge B_{2}
 \right) \,. \nn
\ea
For later purposes we note that in type IIA supergravity there are no self-dual fluxes, so it would be very generally possible to go to a frame where there are no space filling fluxes.

\subsection*{10D Type IIB supergravity}
\label{tobe}

The Ramond-Ramond gauge potentials appearing in Type IIB supergravity are $C$, $C_{\ssM \ssN}$ and $C_{\ssM \ssN \ssP \ssQ}$, and the Einstein-frame lagrangian density for the bosonic sector of the theory is given by
\bea \label{EFlagr}
 \cL &=& - \frac{1}{2\kappa_{10}^2} \sqrt{-\hat g_{10}} \; \left[ \hat g^{\ssM \ssN} \left(
 \hat \cR_{\ssM \ssN} + \frac{\partial_\ssM \tau \partial_\ssN \overline \tau}{2
  \left(\hbox{Im} \, \tau \right)^2} \right)
  + \frac{1}{12 \, \hbox{Im} \, \tau} \, G_{\ssM \ssP \ssR} \overline{G}^{\ssM \ssP \ssR} \right. \nn\\
 && \qquad \left.
 + \frac{1}{480} \, \widetilde{F}_{\ssM \ssP \ssR \ssT \ssV}
 \widetilde{F}^{\ssM \ssP \ssR \ssT \ssV} \right]
  - \frac{i}{8\kappa_{10}^2} \frac{ C_{(4)} \wedge
 G_{(3)} \wedge \overline G_{(3)}}{\hbox{Im} \, \tau}  \,.
\eea
Here the complex fields $\tau$ and $G_{(3)}$ are defined by
\be \label{tauGdefs}
 \tau := C + i \, e^{-\phi}
 \quad \hbox{and} \quad
 G_{(3)} := F_{(3)} - \tau \, H_{(3)} \,,
\ee
where $F_{(k+1)} := \exd C_{(k)}$, $H_{(3)} := \exd B_{(2)}$, and the $\widetilde F$'s are defined by
\be \label{tildeFdefs}
 \widetilde F_{(3)} := F_{(3)} - C H_{(3)}
 \quad \hbox{and} \quad
 \widetilde F_{(5)} = * \widetilde F_{(5)} := F_{(5)} - \frac12 \,
 C_{(2)} \wedge H_{(3)} + \frac12 \, B_{(2)} \wedge F_{(3)} \,.
\ee

This lagrangian scales as $\cL^{\rm IIB} \to s^2 \, \cL^{\rm IIB}$ if the fields are scaled as follows
\be \label{scalingdefIIB}
 e^{-\phi} \to s \, e^{-\phi} \,, \quad
  \hat g_{\ssM \ssN} \to \sqrt{s} \; \hat g_{\ssM \ssN} \,, \quad
 C_{(k)} \to s \, C_{(k)} \,, \quad
 B_{(2)} \to B_{(2)} \,,
\ee
and so becomes a total derivative when evaluated on shell. As computed in Appendix \ref{app:onshell}, the total derivative turns out to be
\ba
 S^{\rm IIB}_{\rm{on-shell}} &=& - {1 \over 8 \kappa^{2}_{10}}
 \int \exd \bigg[    {C}_{2} \wedge   e^{\phi}
 *\widetilde {F}_{3}
 +  {B}_{2}
 \wedge  \Bigl( e^{-\phi}*{H}_{3}  - C_{0} e^{\phi}  * \widetilde{F}_{3} \Bigr)  \nn\\
 && \qquad\qquad\qquad \qquad\qquad \qquad  +
  C_4 \wedge C_2 \wedge H_3 -
   C_4 \wedge F_3 \wedge B_2 \bigg] \,.
\ea

Why should we care when the bulk contribution on the right-hand-side of eq.~\pref{higher} is a total derivative? We care precisely because the bulk fields are generically singular at the specific points in the $n$ compact dimensions where the sources are located. To deal with this singularity, as well as any singularities coming from $S_{\rm source}$, we imagine surrounding these objects in the transverse dimensions by a `Gaussian pillbox' at a small proper distance from the source. This removes the singularity at the source at the expense of introducing a new boundary on the Gaussian pillbox.

When the bulk contribution to the right-hand-side of eq.~\pref{higher} is a total derivative, its integral depends only on the near-source limit of the back-reacted bulk fields at the pillbox. And these boundary conditions, in turn, are related to the physical properties of the source at $y_c^m$ allowing them to be combined with the $S_{\rm source}$ terms in a general way, as the next section discusses in more detail.

The upshot is that although explicitly finding the back-reacted bulk solution for a given source is very difficult, when the curvature depends only on a total derivative most of the details of these solutions are not important. It is only their near-brane boundary conditions that play any role in fixing the on-source curvature, $R$.

\subsection{Step 3: Sources and singularities}
\label{secpb}

The final step is to relate more precisely the boundary contributions to the bulk integrals encountered above to the properties of the source action, $S_{\rm source}$. As we now see, this allows contribution $II$ to be related to contribution $III$ in eq.~\pref{higher}, with the result that they cancel for codimension-two sources.

The trick when doing so is to deal properly with the singularity of the bulk configurations near the sources. We follow a strategy familiar from experience with the Coulomb singularity of electrostatics: we surround the sources with small `Gaussian pillboxes,' and replace the singular extrapolation into the pillbox interior with an appropriate set of boundary conditions on the surface of the box. In this way the singular physics of a point charge is finessed into a finite flux through an arbitrary, but small, surface enclosing the charge.

Of course, this is only a useful construction if the size of the charge distribution is much smaller than the distances of interest for predicting the resulting electric field. If the box is too small compared with the charge distribution inside, the real charge distribution inside cannot be approximated by a point source with the same total charge. A similar problem arises if the box is too large compared with the scales over which the electric fields are to be computed. The construction is useful if a sufficiently large hierarchy exists between the size of the source and the distances of interest for the resulting electric fields.

The same is possible for gravitating systems, provided the physical size of the source is much smaller than the distance over which the gravitational field extends (like the size of any extra dimensions). To accomplish this in the present context \cite{Cod2Matching, BBvN}, we excise a small $D$-dimensional spacetime volume from around each source, and instead specify the boundary conditions on boundary to this small volume.

In the spirit of replacing a real charge distribution by an equivalent point charge, the boundary conditions are specified by doing so for a simple source distribution that shares the same energy. This is most simply done by imagining the source energy density to be distributed on the boundary of the pillbox itself, with the pillbox interior filled in with a smooth field configuration. Such a simple-minded procedure suffices to capture the long-distance physics of a generic real distribution if the pillbox is sufficiently small, with the size of the actual source of interest being much smaller still.

Formally this is done by specifying a $(D-1)$-dimensional codimension-one boundary action, $\widetilde S_{\rm bdy}$, on the pillbox surface, together with a smooth solution describing the pillbox interior. This construction allows boundary conditions to be inferred using standard methods involving the Israel junction conditions \cite{IJC}, which relate $\widetilde S_{\rm bdy}$ to the jump in bulk-field derivatives between inside and outside of the pillbox.

Once these junction conditions are found, a new point of view is possible for which the pillbox is regarded as an honest-to-God boundary of the bulk geometry, without reference to the pillbox interior. In this case one defines a new boundary action for the pillbox, $S_{\rm bdy}$, which is defined by the condition that its derivatives determine the near-source radial derivatives of the fields exterior to the pillbox. In general $S_{\rm bdy}$ differs from $\widetilde S_{\rm bdy}$ because it must now also include any effects that used to be generated by the now non-existent interior geometry. $S_{\rm bdy}$ also includes the Gibbons-Hawking action \cite{GH} for gravity on the boundary, both of the interior and exterior regions:
\be
 S_{\rm bdy} := \widetilde S_{\rm bdy} + S_{\ssG \ssH +} + S_{\ssG \ssH -} + S_{\rm int} \,,
\ee
with
\begin{equation}
 S_{\ssG \ssH} = \frac{1}{\kappa_\ssD^2} \int \exd^{D-1} x  \sqrt{-\gamma} \; K \,,
\end{equation}
and $K = g^{ij} K_{ij}$, where $K_{ij}$ is the extrinsic curvature of the boundary and $\gamma_{ij}$ the induced metric. The subscript $\pm$ for $S_{\ssG \ssH \pm}$ indicates whether the extrinsic curvature is to be computed just inside or just outside of the codimension-one pillbox boundary. The Gibbons-Hawking action is required in the presence of boundaries to make the variation of the Einstein action well-posed. Finally, $S_{\rm int}$ describes the `bulk' action describing the interior geometry, whose details are not important in what follows when the pillbox is sufficiently small.

In the limit of a vanishingly small pillbox, these codimension-one actions can be compactified into corresponding higher-codimension actions. We define $\widetilde S_{\rm source}$ to be the result obtained from $\widetilde S_{\rm bdy}$ in this way, but it is the dimensional reduction of $S_{\rm bdy}$ that compactifies to the $d$-dimensional source action, $S_{\rm source}$, used in previous sections.

This procedure has been worked through in detail for scalar-tensor-Maxwell theories with codimension-two sources in $D = d+2$ dimensions \cite{Cod2Matching}, to which we now specialize. The resulting boundary conditions were then checked for $D7$-brane sources in Type IIB supergravity in 10 dimensions, for which the bulk and source actions are explicitly known, as are a broad class of solutions to the bulk field equations \cite{scs}. In all cases the solutions and actions satisfy the boundary conditions inferred using this simple-minded pillbox construction \cite{BBvN}.

For the present purposes it turns out that we need only the boundary conditions for the metric. Using the Israel junction conditions to relate an assumed smooth interior geometry for the pillbox to the geometry outside, one finds the following junction conditions, expressed in terms of the codimension-one action, $\widetilde S_{\rm bdy}$, of the codimension-one source:\footnote{The difference in signs compared to \cite{BBvN} arises from the choice of unit normal. Here, $K$ is defined with respected to the outward pointing normal, to agree with the convention for the Gibbons-Hawking term.}
\begin{equation} \label{cod1metricmatching}
  \frac{1}{2\kappa_\ssD^2} \, \sqrt{-\hat g_\ssD} \;
   \left( K^{ij} - K g^{ij} \right) - \hbox{(int)}^{ij} =
   \; \frac{\delta \widetilde S_{\rm bdy}}{\delta \hat g_{ij}} \,.
\end{equation}
This expression adopts coordinates near the pillbox for which $\rho$ denotes radial proper distance away from the source, which is located at $\rho = 0$. The pillbox boundary lies on a surface of fixed, small $\rho$, for which $K_{ij}$ is the extrinsic curvature of the fixed-$\rho$ surface, for which the local coordinates are $\{ x^i \} = \{ x^\mu, \theta \}$, with $i = 0, 1, \cdots, d$ where $d = D-2$ and $\theta$ is an angular coordinate that runs from 0 to $2\pi$ as one encircles the source. Finally, `(int)${}^{ij}$' denotes the same result evaluated for the smooth interior geometry, for which $\rho = 0$ is nonsingular.

As mentioned earlier, there are two equivalent ways to read eq.~\pref{cod1metricmatching}. The first is the way it was initially derived: where $\widetilde S_{\rm bdy}$ represents only the action of the boundary, and the interior region of the brane is matched onto the exterior one through eq.~\pref{cod1metricmatching}. The other viewpoint is that the pillbox is considered the actual boundary of spacetime, and the `interior' of the branes is excised entirely. In this point of view, the properties of the interior solutions are encoded in the boundary action, $S_{\rm bdy}$:
\begin{equation} \label{cod1metricmatching2}
  \frac{1}{2\kappa_\ssD^2} \, \sqrt{-\hat g_\ssD} \;
   \left( K^{ij} - K g^{ij} \right)  =
   \frac{\delta \widetilde S_{\rm bdy}}{\delta \hat g_{ij}}
   +  \hbox{(int)}^{ij} = \frac{\delta S_{\rm bdy}}{\delta
   \hat g_{ij}} \,.
\end{equation}

In the limit of a very small pillbox, these conditions dimensionally reduce to conditions that only refer to the codimension-two action.
\begin{equation}
   \lim_{\rho \to 0} \oint_{x_b} \exd \theta \;
    \left[ \frac{1}{2\kappa_\ssD^2} \, \sqrt{-\hat g} \;
   \left( K^{ij} - K \hat g^{ij} \right)
   - \hbox{(int)}{}^{ij}\right] =
   \; \frac{\delta \widetilde S_{\rm source}}{\delta \hat g_{ij}} \,,
\end{equation}
where the integration is about a small circle of proper radius $\rho$ encircling the brane position at $\rho = 0$, and $N_\ssM$ is the unit normal pointing towards the brane ($N_\ssM\exd x^\ssM=-\exd\rho$).

The upshot is that source-bulk matching relates the asymptotic, near-source radial derivatives of the bulk fields to the properties of the source action. In what follows, an important role is played by the function, $U_{\rm source}$, that controls the codimension-two boundary condition for the warp factor, $W$,
\ba \label{finalmatching}
 \frac{ d}{\kappa_\ssD^2} \lim_{\rho \to 0} \oint \exd \theta
 \sqrt{-\hat g_\ssD} \; N^\ssM\pd_\ssM W &=& 2\frac\pd{\pd g_{\theta\theta}}\left[\sqrt{-g_d}\;
 \widetilde \cL_{\rm source}\right] \nn\\
 &:=& d\sqrt{-g_d} \; U_{\rm source}\,,
\ea
where the last equality defines $U_{\rm source}$, and $\widetilde \cL_{\rm source}$ is the codimension-two lagrange density
\be
 \widetilde S_{\rm source}= \int \exd^d x \sqrt{-\hat g_d} \;
 \widetilde \cL_{\rm source}\,.
\ee

The function $U_{\rm source}$ is important\footnote{Although determination of $U_{\rm source}$ appears to require knowing how $S_{\rm source}$ depends on $g_{\theta\theta}$, this is actually not necessary because the it is related \cite{BBvN} by an identity --- the `Hamiltonian' constraint for evolution in the $\rho$ direction, since this relates the first derivatives of bulk fields with respect to $\rho$ --- to the easily computed derivatives $\delta S_{\rm source}/\delta \phi^a$ and $\delta S_{\rm source}/\delta g_{\mu\nu}$.} for other reasons, besides its above role in controlling the asymptotic behaviour of the warp factor. As we show below, for codimension-two sources $U_{\rm source}$ turns out also to be the Lagrange density of the full action, $S_{\rm source}$ \cite{Cod2Matching, BBvN}. It turns out that $U_{\rm source}$ is generically non-negative, and this is related to the general property (described below) that the bulk field equations dictate that $W$ does not increase as one approaches a codimension-two source.

\subsubsection*{Implications for the on-source curvature}

We now show how the above matching conditions imply a dramatic cancelation in our key formula, eq.~\pref{higher}. In particular, after using Gauss' law to rewrite total derivatives in terms of surface terms at the position of the Gaussian pillboxes surrounding the sources, followed by eq.~\pref{finalmatching}, one of the terms on the right-hand-side of eq.~\pref{higher} can be written:
\ba
\label{cancelterms}
 { 1 \over 2 \kappa_{\ssD}^{2} }
  \int \exd^{d} x \sqrt{-{g}_{d}} \; \int \exd^{2} y \sqrt{\tilde{g}_{2}} \; \tilde{\nabla}^2  e^{dW}
 &=& { d \over 2 \kappa_{\ssD}^{2} }
  \int \exd^{d} x \sqrt{-{g}_{d}} \; \oint \exd \theta \sqrt{\tilde{g}_{2}} \; (N\cdot \tilde\nabla W)  e^{dW}  \nn\\
 &=&  { d \over 2 }
  \int \exd^{d} x \sqrt{-{g}_{d}} \; U_{\rm source} \,.
\ea
We wish to compare this with another term on the right-hand-side of eq.~\pref{higher},
\be
 \int \exd^d x \; \hat g_{\mu\nu}
  \left( \frac{\delta S_{\rm source}}
 {\delta \hat g_{\mu\nu}} \right)
 = \lim_{\rho \to 0} \int \exd^{d+1} x \; \hat g_{\mu\nu}
  \left( \frac{\delta S_{\rm bdy}}
 {\delta \hat g_{\mu\nu}} \right) \,.
\ee
To evaluate this we use the matching condition, eq.~\pref{cod1metricmatching}, which implies
\be \label{cod1metricmatchingtr}
   \int \exd^{d+1} x \; \hat g_{ij} \frac{\delta \widetilde
   S_{\rm bdy}}{\delta \hat g_{ij}}
  = - \frac{d}{2\kappa_\ssD^2} \int \exd^{d+1} x
   \sqrt{-\hat g_\ssD} \; \Bigl[  K - \hbox{(int)} \Bigr]
   = - \frac{d}{2} \, \Bigl( S_{\ssG\ssH+} + S_{\ssG\ssH-}
   \Bigr) \,,
\ee
to rewrite $S_{\rm bdy}$ as follows:
\ba \label{bdybdytilde}
 S_{\rm bdy} &=& \widetilde S_{\rm bdy} + S_{\ssG\ssH+}
 + S_{\ssG\ssH-}\nn\\
 &=& \widetilde S_{\rm bdy} - \frac{2}{d} \int \exd^{d+1} x \; \hat g_{ij} \, \frac{\delta \widetilde S_{\rm bdy}}{ \delta \hat g_{ij}} \nn\\
 &=& \widetilde S_{\rm bdy} - \frac{2}{d} \int \exd^{d+1} x \; \left( \hat g_{\mu\nu} \, \frac{\delta \widetilde S_{\rm bdy}}{ \delta \hat g_{\mu\nu}} + \hat g_{\theta\theta} \, \frac{\delta \widetilde S_{\rm bdy}}{ \delta \hat g_{\theta\theta}} \right)\,,
\ea
Now, our interest is in maximally symmetric configurations with no space-filling fluxes, for which
\be
 \widetilde S_{\rm bdy} = \int \exd^{d+1} x \sqrt{- \hat g_\ssD}
 \; \widetilde \cL_{\rm bdy} \,,
\ee
and $\widetilde \cL_{\rm bdy}$ does not depend on curvatures. In this case $\delta \widetilde S_{\rm bdy}/\delta \hat g_{\mu\nu} = \frac12 \, \sqrt{- \hat g_\ssD} \; \widetilde \cL_{\rm bdy} \, \hat g^{\mu\nu}$.  Using this in eq.~\pref{bdybdytilde} gives
\be
 S_{\rm source} = \lim_{\rho \to 0} S_{\rm bdy}
 = - \frac{2}{d} \lim_{\rho \to 0} \int \exd^{d+1} x
 \; \hat g_{\theta\theta} \, \frac{\delta \widetilde S_{\rm bdy}}{ \delta \hat g_{\theta\theta}}
 =  -\int \exd^{d} x \sqrt{- g_{d}}
 \; U_{\rm source} \,,
\ee
where the last equality uses eq.~\pref{finalmatching}. This leads finally to our desired expression:
\be \label{cancelee2}
 \int \exd^d x \; \hat g_{\mu\nu}
  \left( \frac{\delta S_{\rm source}}
 {\delta \hat g_{\mu\nu}} \right)
 = - \frac{d}{2} \int \exd^{d} x \sqrt{-g_d} \; U_{\rm source} \,.
\ee
As claimed, from eqs.~\pref{cancelterms} and \pref{cancelee2} we see that the codimension-two matching conditions ensure the cancelation of two of the terms on the right-hand-side of eq.~\pref{higher},
\be
\label{canceltermsfinal}
 { 1 \over 2 \kappa_{\ssD}^{2} }
  \int \exd^{d} x \sqrt{-{g}_{d}} \; \int \exd^{2} y \sqrt{\tilde{g}_{2}} \; \tilde{\nabla}^2  e^{dW}
  + \int \exd^d x \; \hat g_{\mu\nu}
  \left( \frac{\delta S_{\rm source}}
 {\delta \hat g_{\mu\nu}} \right) = 0 \,,
\ee
leaving
\ba
\label{higherlast}
 -{ 1 \over 2 \kappa_{d}^{2} }
 \int \exd^{d} x \sqrt{-{g}_{d}} \; R
 &=& {d \over 2 } \; S_{\rm on-shell}
   + { 1 \over 2 \kappa_{\ssD}^{2} }
 \int \exd^{\ssD}x  \sqrt{-\hat{g}_{\ssD}} \;
 \hat{g}_{\mu \nu}{ \partial
 {\cal{L} }^{\ssD}_{\rm matter} \over
 \partial \hat{g}_{\mu \nu} } \nn\\
 &=& {d \over 2 } \; S_{\rm on-shell}
  \,,
\ea
with the second line following because we already assumed there to be no space-filling fluxes. This, together with the earlier expressions that give $S_{\rm on-shell}$ as a total derivative, are our main results.

\section{Example: the axio-dilaton and 10D Type IIB supergravity}

Our goal in this section is to illustrate the generality of the result, eq.~\pref{higherlast}, obtained at the end of the last section. We use eq.~\pref{higherlast} to show that the on-source curvature vanishes for F-theory axio-dilaton compactifications of 10D Type IIB supergravity with arbitrary codimension-two sources, generalizing a known result when the sources are supersymmetric \cite{ftheory}. Although this example corresponds to the choices $d = 8$ and $n = 2$, --- with only the metric, $g_{\ssM\ssN}$, and the axio-dilaton, $\tau = C + i \, e^\phi$, (and no other fluxes) in play, in what follows we work instead with general $d$.

This choice is made for three reasons. First, because it includes a broad class of explicitly known solutions \cite{scs} with explicit sources: $D7$- and $O7$-planes, as well as various kinds of $(p,q)$-branes. Second, because the absence of bulk fluxes ensures that the right-hand-side of eq.~\pref{higher} is particularly simple (and is a total derivative). Third, the $d$-dimensional sources in this case have codimension two, which is one of the few situations for which matching conditions relating near-source asymptotics to physical properties of the source are explicitly worked out \cite{Cod2Matching}. In particular, they have been tested explicitly \cite{BBvN} for the solutions of ref.~\cite{scs} with $D7$-brane sources --- and implicitly, using $SL(2,R)$ invariance, for $(p,q)$-brane sources as well.

\subsection{Bulk equations}

The Einstein frame action for the Einstein-axio-dilaton system in 10D Type IIB supergravity is $S = S_\ssB + S_{\rm source}$, where
\begin{equation} \label{ftheoryaction}
 S_\ssB = - \frac{1}{2\kappa^2}
 \int \exd^{10}x \sqrt{- \hat g} \; \hat g^{\ssM \ssN}
 \left[ \hat \cR_{\ssM \ssN}
 + \frac{\partial_\ssM \overline\tau
 \,\partial_\ssN \tau}{2\, (\hbox{Im}\, \tau)^2} \right]
 \,.
\end{equation}
This is invariant under PSL(2,$R$) transformations
\be
 \tau \to \frac{a \tau + b}{c \tau + d} \,,
\ee
with the real parameters $a$ through $d$ satisfying $a\,d-b\,c = 1$. The scaling symmetry boils down in this case to $\tau \to s \, \tau$ and $\hat g_{\ssM \ssN} \to \sqrt{s} \; \hat g_{\ssM \ssN}$, under which $S_\ssB \to s^2 \, S_\ssB$.

The Einstein field equations for this action are
\be
 \hat{\cR}_{\ssM \ssN} + \frac{1}{4({\rm Im} \, \tau)^{2}} \left( \partial_{\ssM} \bar{\tau} \partial_{\ssN} \tau
 + \partial_{\ssN} \bar{\tau} \partial_{\ssM} \tau \right)
  = \hbox{(source terms)} \,,
\ee
whose trace with $\hat g^{\ssM \ssN}$ ensures that $S_{\rm on-shell} = 0$ (for all $D$). The axio-dilaton equation is, similarly
\be
 -i \hat\nabla^2 \tau + \frac{\partial^\ssM \tau
 \partial_\ssM \tau}{{\rm Im} \, \tau} =
 \hbox{(source terms)} \,.
\ee

As ever, the solutions of interest have geometry
\be
 \exd {\hat{s}}^{2} = \hat{g}_{\ssM \ssN} \, \exd x^{\ssM}
 \exd x^{\ssN} = e^{2W} \, {g}_{\mu \nu} \, \exd x^{\mu} \exd x^{\nu} + \tilde{g}_{mn} \, \exd y^{m} \exd y^{n} \,,
\ee
where $g_{\mu\nu}(x)$ is a $d$-dimensional maximally symmetric Minkowski-signature metric, and $W(y)$, $\tau(y)$ and $\tilde g_{mn}(y)$ depend only on the other $n$ compact directions. We temporarily keep the variables $d$ and $n$ general, although at the end we specialize to our real interest in this section: $n = 2$ (and $D = 10$ and $d = 8$, though this is less crucial).

For general $d$ and $n$ the Ricci tensors satisfy
\ba
 \hat{\cR}_{\mu\nu} &=& R_{\mu\nu} + \Bigl(
 \tilde{\nabla}^{2} W + d \, \tilde g^{mn}
 \partial_m W \, \partial_n W \Bigr) e^{2W}
  g_{\mu\nu} \nn\\
   &=& R_{\mu\nu} + \frac{1}{d} \, e^{(2-d)W}
  \left(  \tilde{\nabla}^{2} e^{dW} \right) g_{\mu\nu} \nn\\
 \hbox{and} \quad
 \hat{g}^{mn} \hat{\cR}_{mn} &=& \tilde R + d \Bigl(
 \tilde \nabla^2 W + \tilde g^{mn} \partial_m W
 \partial_n W \Bigr)
 =\tilde R + d \, e^{-W} \, \tilde{\nabla}^{2}e^{W} \,,
\ea
and so the $(\mu\nu)$ Einstein equations, $\hat \cR_{\mu \nu} = 0$, boil down to
\be \label{einsmunu}
 R \, e^{-2W} + e^{-dW} \tilde{\nabla}^{2} e^{dW} =
 \hbox{(source terms)} \,,
\ee
while the $n$-dimensional trace of the remaining Einstein equations becomes
\be \label{einsmn}
  \tilde R  +  d \, e^{-W} \tilde{\nabla}^{2} e^W
 + \frac{ \tilde g^{mn} \partial_m \tau
 \partial_n \bar{\tau}}{2
 ({\rm Im} \, \tau)^{2}}  = \hbox{(source terms)} \,.
\ee
A broad class of unwarped solutions to these equations are known \cite{scs}, and reviewed in Appendix \ref{app:ftheorysolns}.

\subsubsection*{Codimension-two sources}

Because source-bulk matching is best understood for codimension-two, we specialize now to the case $n = 2$, in which case several things simplify.

First, the trace leading to the last equation carries no loss of information, and so the full set of Einstein equations become completely equivalent to eqs.~\pref{einsmunu} and \pref{einsmn}. Second, it becomes convenient to use complex coordinates, $z := x^{8}+ix^{9} = y^1 + i y^2$, and write the compact metric in conformally flat form
\be
 \tilde{g}_{mn} \,\exd x^{m} \exd x^{n}
 = e^{2C} \, \exd z \, \exd \bar{z}
 = \exd \rho^2 + e^{2B} \, \exd \theta^2 \,.
\ee
With these choices $\tilde{\nabla}^{2} f = e^{-2C} \, \delta^{mn} \partial_{m} \partial_{n} f = 4 \, e^{-2C} \, \partial \bar{\partial} f$, for any scalar field $f$, and
the scalar curvature becomes $\tilde R = 2\, \tilde{\nabla}^{2} C$.

The Einstein equations simplify to
\ba \label{Einsteincomplex}
 \frac14 \, R \, e^{2C} + e^{-dW} \partial
 \bar \partial e^{dW} &=& 0 \nn\\
 2 \, \partial \bar \partial C + d\, e^{-W} \partial \bar
 \partial e^W - \frac{ (\partial \tau \, \bar \partial
 \bar \tau + \partial \bar \tau \, \bar \partial \tau)}{
 (\tau - \bar \tau)^2} &=& 0 \,,
\ea
while the axio-dilaton equation of motion becomes independent of $C$:
\be \label{axiocomplex}
 \partial \bar\partial \,\tau + \frac{d}2 \, ( \partial W
 \bar \partial \tau + \bar \partial W \partial \tau)
 + \frac{2 \,
 \partial\tau \bar{\partial} \tau}{\bar{\tau} - \tau}
  = 0 \,.
\ee

Finally, we identify the contributions on the right-hand-side of eq.~\pref{higher} for this example. Since there are no space-filling fluxes and the on-shell action vanishes, eq.~\pref{higherlast} for this example reduces to
\be
\label{higherz}
  R = \frac{d}{2} \, S_{\rm on-shell}
 =0  \,.
\ee
Since $R=0$, eqn.~\pref{Einsteincomplex} implies that $e^{dW}$ is the real part of a holomorphic function.

Notice that if we had not included the source term, our conventions are such that the warping term contributes an AdS sign if $N \cdot \partial W < 0$; {\em i.e.} $W$ {\em decreases} towards the boundary. As we show below, the explicit asymptotic form for the bulk solution near the sources can be found in general, and for a codimension-two source situated at $\rho = 0$ (where $\rho$ denotes proper radius) has the form $e^W \propto \rho^\omega$ with $\omega \ge 0$, in agreement with the AdS sign found in the no-go results \cite{GG, MN, Wox, StW}.

\subsection{Near-source Kasner solutions}
\label{app:nearbranelimit}

To find asymptotic solutions in the vicinity of a source it is convenient to use an orthogonal coordinate system including proper distance $\rho$. We therefore take the following {\em ansatz} for the metric and dilaton
\ba
 \hat{\exd s}^2 &=&  \exd\rho^2 + \cA\rho^{2\alpha} \exd\theta^2+ \cB \rho^{2\omega} \, g_{\mu\nu}
 \exd x^\mu \exd x^\nu \nn\\
 \tau&=&k\theta+i\cF \rho^{-q}\,,
\ea
where $\cA = a_0 + a_1 \ln \rho$, $\cB = b_0 + b_1 \ln \rho$ and $\cF = f_0 + f_1 \ln \rho$. This form captures, in particular, the asymptotic form of the known unwarped solutions described in Appendix \ref{app:ftheorysolns}. Since the quantity $b_1$ first arises in the field equations at subdominant order as $\rho \to 0$, we initially neglect it here.

Given this choice, and keeping only the most singular part as $\rho \to 0$, the dilaton equation becomes
\ba
 &&\rho^{-q-2} \left[ (\alpha+d\omega-1) (f_1-qf_0-qf_1\ln\rho)
 - \frac{f_1^2}{f_0+f_1\ln\rho} \right]\nn\\
 &&\qquad\qquad+ \rho^{-q-2} \left[ \frac{a_1}2 \frac{f_1-qf_0-qf_1 \ln\rho}{a_0+a_1\ln\rho} + \frac{k^2\rho^{2q+2-2\alpha}}{(a_0+a_1\ln\rho)
 (f_0+f_1\ln\rho)}\right]=0 \,.
\ea
We keep the variable $d$ general here, although our Type IIB application is to $d = 8$. The ($\rho\rho$) Einstein equation similarly is
\ba
 0&=&\frac1{\rho^2} \left[ \alpha(\alpha-1)+d\omega(\omega-1)+\frac12q^2 \right] + \frac1{\rho^2}\left[\frac{a_1(2\alpha-1)}{2(a_0+a_1\ln\rho)} -\frac{qf_1}{f_0+f_1\ln\rho}  \right]\nn\\
 &&\qquad\qquad + \frac1{\rho^2}\left[ \frac{f_1^2}{2(f_o+f_1\ln\rho)^2}-\frac{a_1^2}{4(a_0+a_1\ln\rho)^2} \right]\,,
\ea
while the ($\theta\theta$) equation gives
\be
 \frac{g_{\theta\theta}}{\rho^2}\left[\alpha(\alpha+d\omega-1)+\frac{a_1(2\alpha+d\omega-1 )}{2(a_0+a_1\ln\rho)}-\frac14\frac{a_1^2}{(a_0+a_1\ln\rho)^2} +\frac{k^2\rho^{2q+2-2\alpha}}{4(a_0+a_1\ln\rho)(f_0+f_1\ln\rho)^2}\right]=0 \,.
\ee

To leading approximation the most singular part of these equations as $\rho \to 0$ is solved --- up to terms of relative order $1/\ln\rho$ or more --- if the powers satisfy the two `Kasner' conditions,
\ba
 \alpha+d\omega-1&=&0\nn\\
 \alpha(\alpha-1)+d \, \omega(\omega-1) + \frac{ q^2}{2}&=&0 \,.
\ea
Using the first of these to simplify the latter allows it to be written
\be
 \alpha^2+d \, \omega^2+\frac{q^2}{2}=1 \,.
\ee

This result holds if terms that depend on $k$ are suppressed, which is true if the condition $q+1>\alpha$ is satisfied. In the case of interest, with $d=8$, $\alpha$ can be eliminated from the Kasner conditions to give
\be
 72 \, \omega^2 - 16 \, \omega+\frac{q^2}{2}=0\,,
\ee
with solutions
\be
 \omega=\frac19 \left( 1\pm\sqrt{1-\frac{9 q^2}{16}} \right)\,.
\ee
This shows that the only real solutions have $\omega \ge 0$, and consequently $\alpha \le 1$. The limiting case with $q=\omega=0$ and $\alpha=1$ corresponds to a conical singularity at the brane position. Hence positive $q$ is sufficient to have the Kasner condition satisfy the leading terms in the field equations near $\rho = 0$, with additional contributions of order $1/\ln\rho$ and smaller.

Notice in particular that because $\omega \ge 0$, the warp factor always either goes to zero or to a finite value when approaching a source. This ensures that the warping contribution to eq.~\pref{higher} is never of the de Sitter sign.

We can now consider what happens if we do not neglect the logarithm, $b_1 \ln \rho$, in the warping. In this case
\be
 \hat g_{\mu\nu}=\rho^{2\omega}(W_0+W_1\ln\rho)g_{\mu\nu} \,.
\ee
In the dilaton equation, we get the additional (suppressed) terms
\be
 ...+\rho^{-q-2}\left[ \frac{W_1}2\frac{f_1-q f_0-q f_1\ln\rho}{W_0+W_1\ln\rho}=0 \right]\,.
\ee
In the ($\rho\rho$) Einstein equation this gives
\be
 ...+\frac1{\rho^2}\left[\frac{\omega}{W_0+W_1\ln\rho}-\frac12\frac{W_1}{W_0+W_1\ln\rho} -\frac14\frac{W_1^2}{(W_0+W_1\ln\rho)^2}\right]\,,
\ee
and finally for ($\theta\theta$)
\be
 ...-\frac{g_{\theta\theta}}{\rho^2}\left[\frac d2\frac{\alpha W_1}{W_0+W_1\ln\rho}-\frac d4\frac{a_1W_1}{(a_0+a_1\ln\rho)(W_0+W_1\ln\rho)} \right]\,.
\ee
From this we see that a log-term in $W$ only modifies the field equations at a suppressed $1/\ln\rho$ level.

\section{Conclusions}

In summary, in this paper we examine solutions to extra-dimensional field equations for geometries of the form of eq.~\pref{metricform}, with maximal symmetry in the noncompact dimensions. We ask what features of a solution control the curvature in the maximally symmetric, noncompact dimensions.

Our main result is given by eq.~\pref{higher}, which gives the noncompact curvature scalar as a sum of four terms: $R \propto I + II + III + IV$. Here $I$ corresponds to the bulk action evaluated at the appropriate back-reacted solution; $II$ denotes an integral over a total derivative involving the warp factor (whose sign is usually definite, and not de Sitter-like); $III$ denotes the direct contribution of the actions of any localized sources; and $IV$ denotes a term which vanishes for solutions that are maximally symmetric in the noncompact dimensions, in the absence of space-filling fluxes.

Our main new result is to show, for codimension-two sources, that the boundary conditions that must be satisfied near the sources relate the near-source asymptotics of the bulk fields in such a way that the contributions $II$ and $III$ precisely cancel.

In these circumstances eq.~\pref{higher} degenerates down to eq.~\pref{higherlast}, which relates the curvature completely to the on-shell bulk action. Remarkably, it is very often true that this on-shell action is also a total derivative. A sufficient condition for this turns out to be the existence of a rigid scale invariance of the classical equations of motion \cite{WBW6D}, which in particular is present for most higher-dimensional supergravity theories of general interest. When $S_{\rm on-shell}$ is the integral of a total derivative, the curvature of the noncompact dimensions is completely determined by the asymptotic form of a particular combination of bulk fields near any sources that are distributed around the extra dimensions.

These arguments have two main implications. First, they show (at least for codimension-two sources) that source back-reaction and the source actions cannot be neglected when seeking de Sitter solutions. But they also show that all of the details of the complete back-reacted solution are not required; it often suffices to know the asymptotic behaviour of the bulk fields in the near-source limit.

We explicitly derive which bulk fields play this role for 11D supergravity and 10D Type IIA and Type IIB supergravity, and we hope soon to have results to report on new kinds of explicit extra-dimensional de Sitter solutions that can exploit the results we present here.

\section*{Acknowledgements}

We thank Ross Diener for helpful discussions, and the Abdus Salam International Centre for Theoretical Physics and Perimeter Institute for facilitating our collaboration on this project. CB also thanks the Niels Bohr International Academy for its hospitality in the final stages of this work. AAN would like to acknowledge the Abdus Salam ICTP and the Cambridge Commonwealth Trust for financial support. CB and LvN's research is supported in part by funds from the Natural Sciences and Engineering Research Council (NSERC) of Canada. Research at the Perimeter Institute is supported in part by the Government of Canada through NSERC and by the Province of Ontario through MEDT. AM is funded by the EU under the Seventh Framework Programme (FP7) and the University of Cambridge.

\appendix

\section{Curvature and fluxes for simple Freund-Rubin examples}
\label{exam}

In this appendix we review several familiar Freund-Rubin $AdS_{d}\times S_{p}$ solutions to higher-dimensional supergravity, where $d+p=D$. We do so in order to explore how space-filling fluxes show up in eq.~\pref{higher} of the main text.

\subsection*{Freund-Rubin solutions}

Consider solutions to the field equations for the action
\be
 S= - \frac{1}{2 \kappa_{\ssD}^{2}} \int \exd^{D}x \sqrt{-
 g_\ssD} \; \left( \cR + \frac{1}{2 \,p!} \, F^{2} \right) \,.
\ee
For the $p$-form threading a $p$-sphere,  $F_{m_{1}.....m_{p}} = k \, \epsilon_{m_{1}.....m_{p}}$,
Einstein's equations
\be
 \cR_{\ssM \ssN} - \frac{1}{2} \, g_{\ssM \ssN}
  \cR + \frac{1}{2 (p-1)!}
 \, \left( F_{\ssM \ssA \ssB \ssC..}
 F_{\ssN}^{\phantom{\ssM} \ssA \ssB \ssC..}
 - \frac{1}{2p} \, g_{\ssM \ssN} \, F^{2} \right)
 =0 \,,
\ee
yield the solutions that are product spaces,
\be
 \exd s^2 =  g_{\ssM \ssN} \exd x^\ssM \exd x^\ssN =
 g_{\mu\nu} \exd x^\mu \exd x^\nu + \tilde g_{mn} \,
 \exd x^m \exd x^n \,,
\ee
with curvatures
\be
 \tilde R = -\frac{k^{2} p (D-p-1)}{2(D-2)}
 \quad  \hbox{and} \quad
 R = \frac{k^{2}(2p-D)}{2(D-2)} \,.
\ee
Here $\tilde R$ is the Ricci scalar associated with the $p$-sphere metric (which is negative in our conventions), $\tilde g_{mn}$, $R$ is the (positive) Ricci scalar of a $d$-dimensional anti-de Sitter metric, $g_{\mu\nu}$. $\cR_{\ssM \ssN}$ is the Ricci tensor for the full $D$-dimensional metric $g_{\ssM \ssN}$. (In the absence of warping we need not distinguish $\hat g_{\mu\nu}$ from $g_{\mu\nu}$.)

\subsection*{Example: 11D supergravity}

In this section we consider several examples from 11D supergravity that illustrate the equality (\ref{higher}) with and without space-filling fluxes.

Since the Chern-Simons term does not contribute, Freund-Rubin solutions for 11-D supergravity can be obtained using the 4-form field strength, $G_{\ssM\ssN \ssP\ssQ}$, and the following action
\be
   S_{11} = -\frac{1}{2 \kappa_{11}^{2}}
   \int \exd^{11}x \sqrt{-  g_{11}} \;
   \left[ \cR + \frac{1}{2(4!)} \,
   G_{4}^{2} \right] \,.
\ee
There are two natural choices, depending on whether the 4-form flux threads the anti-de Sitter or spherical dimensions.

\medskip\noindent{$AdS_7 \times S_4$}

\medskip\noindent
First consider solutions of the form $AdS_{7} \times S_{4}$, for which the only nonzero components of $G_4$ are along the 4-sphere directions:
\be
 G_{mnpq} = 3n \, \epsilon_{mnpq}
 \quad \hbox{and so} \quad
 G_4^{2} = (9 n^{2}) 4! \,.
\ee
Einstein's equations are
\be \label{sfour}
  \cR_{\ssM \ssN} - \frac{1}{2} \, g_{\ssM\ssN} \, \cR
 + \frac{1}{12} \, \left( G_{\ssM\ssA\ssB\ssC}
 G_{\ssN}^{\phantom{\ssM}\ssA\ssB\ssC}
 - \frac{1}{8} \, g_{\ssM\ssN} \, G_4^{2} \right) = 0\,,
\ee
and so taking the 11-, 7- and 4-dimensional traces of eq.~(\ref{sfour}) one finds
\be
 \cR = -\frac{3 n^{2}}{2} \,,
  \qquad
  R = g^{\mu \nu} \cR_{\mu \nu} =  \frac{21 n^{2}}{2}
 \quad \hbox{and} \quad
 \tilde R = \tilde g^{mn} \tilde R_{mn}
 = -12 n^{2} \,,
\ee
corresponding to $AdS_7 \times S_4$.

One can use these to check eq.~\pref{higher}:
\ba
  -{ 1 \over 2 \kappa_{7}^{2} } \int \exd^{7} x
  \sqrt{-{g}_{7}} \; R &=& -{ 21 \, n^2\over 4 \,
  \kappa_{7}^{2} } \int \exd^{7} x  \nn\\
  \hbox{and} \quad
  S_{\rm on-shell} &=&  - \frac{1}{2\kappa_{11}^2}
  \int \exd^{11} \sqrt{- g_{11}} \; \left[
   -\frac{3 n^{2}}{2} + \frac{(9 n^{2}) 4!}{2(4!)}
   \right] \nn\\
   &=&  - \frac{3n^2}{2 \,\kappa_{11}^2}
  \int \exd^{11} \sqrt{- g_{11}}  \,,\nn
\ea
and so
\be
  -{ 1 \over 2 \kappa_{7}^{2} } \int \exd^{7} x
  \sqrt{-{g}_{7}} \; R = {7 \over 2 }
  \, S_{\rm on-shell} \,,
\ee
as required by (\ref{higher}) for a unwarped solution of maximal symmetry without space filling flux.

\medskip\noindent{$AdS_4 \times S_7$}

\medskip\noindent
Now consider the solution $AdS_{4}\times S_{7}$, which involves a space-filling flux:  $G_{\mu\nu\rho\sigma} = 3m \, \epsilon_{\mu\nu\rho\sigma}$. From Einstein's equations one finds
\be
 \cR = \frac{3 m^2}{2} \,, \quad
 \tilde R = \tilde g^{mn} \tilde R_{mn} = -\frac{21 m^2}{2}
 \quad \hbox{and} \quad
  R = g^{\mu \nu} R_{\mu \nu} = 12m^{2} \,.
\ee
In this case one finds a mismatch between
\be
  -{ 1 \over 2 \kappa_{4}^{2} } \int \exd^{4} x
   \sqrt{-{g}_{4}} \; {{R}} \quad
   {\rm{and}} \quad
   {4 \over 2 } \, S_{\rm on-shell} \,.
\ee
This difference is accounted for by including the flux contribution to $g^{\mu\nu} \partial \cL^{11}/\partial g^{\mu\nu}$, which gives a term of the form of eq.~(\ref{spacef}), as required by eq.~(\ref{higher}).

Alternatively, one can work with a dual Lagrangian containing a kinetic term for the 7-form, $H$, that is dual to $G$:
\be
    S_{\rm {dualized}} = - \frac{1}{2 \kappa_{11}^{2}}
    \int \exd^{11}x \sqrt{-g_{11}} \;
    \left[ \cR + \frac{1}{2(7!)} \, H_{7}^{2}
    \right] \,.
\ee
In this description the seven form threads only internal directions and has no space-filling components, and the dualized action evaluates to
\be
  -{ 1 \over 2 \kappa_{4}^{2} } \int \exd^{4} x
  \sqrt{-{g}_{4}} \, {{R}} =  {4 \over 2 }
  \, S_{\textrm{{\tiny{ on-shell (dualized)}}}} \,.
\ee
Recall for these purposes that although dualization is a symmetry of the equations of motion, it is not a symmetry of the action.

\section{On-shell supergravity actions}
\label{app:onshell}

This appendix explicitly evaluates the total-derivative form for the bosonic sectors of 11D and 10D Type IIA and Type IIB supergravity.

\subsection*{11D supergravity}

For 11D supergravity the bosonic action is
\be
 S^{11} = -\frac{1}{2 \kappa_{11}^{2}}
 \int \exd^{11} x \sqrt{-\hat g_{11}} \; \left[
 \hat \cR + \frac{1}{2(4!)} \, G_{4}^{2} \right]
 - {1 \over 12 \kappa_{11}^{2}}
 \int G_{4} \wedge G_{4} \wedge C_{3} \,.
\ee
This scales as $S^{11} \rightarrow s^{9/2} S^{11}$ when $\hat g_{\ssM \ssN} \to s \hat g_{\ssM \ssN}$ and $C_{\ssM \ssN \ssP} \to s^{3/2} C_{\ssM \ssN \ssP}$.

As argued above, this scaling behaviour implies that the on-shell lagrangian is a total derivative. To show this in detail use the trace of Einstein equation,
\be
 \hat \cR = - \frac{G_{4}^{2}}{6(4!)} \,,
\ee
and the equation of motion for the 3-form potential:
\be
 \exd(*G_{4}) = - \frac{1}{2} \,
 G_{4} \wedge G_{4} \,.
\ee
Together, these two equations give the following expression for the on-shell 11D action:
\be
 S^{11}_{\rm on-shell} = -\frac{1}{6 \kappa_{11}^{2}}
 \int \exd \Bigl(C_{3} \wedge *G_{4} \Bigr) \,.
\ee

\subsection*{10D Type IIA supergravity}

The story for the 10D Type IIA supergravity action is similar. In the string frame this action is the sum of the Neveu-Schwarz, Ramond-Ramond and Chern-Simons sectors,
\be
 S^{\rm IIA} = S_{NS}+S_{RR}+S_{CS} \,,
\ee
where
 \ba
  S_{NS} &=& - \frac{1}{2\kappa_{10}^{2}}
  \int \exd^{10} x \sqrt{-\hat{g}_{10}} \;
  e^{-2\phi} \left[ \hat{\cR} - 4 \, \partial_{\ssM} \phi \partial^{\ssM} \phi + \frac{1}{2(3!)} \,
  H_{3}^{2} \right] \nn\\
 S_{RR} &=& - \frac{1}{2 \kappa_{10}^{2}}
 \int \exd^{10} x \sqrt{-\hat{g}_{10}} \;
 \left[ \frac{1}{2(2!)} \, F_{2}^{2} + \frac{1}{2(4!)} \,
 \tilde{F}_{4}^{2} \right] \\
 S_{CS} &=& - \frac{1}{4 \kappa_{10}^{2}}
 \int B_{2} \wedge F_{4} \wedge F_{4} \,, \nn
\ea
and
\be
 \tilde{F}_{4} = F_{4} + C_{1} \wedge H_{3} \,,
 \quad
 H_{3} = \exd B_{2} \,, \quad
 F_{2} = \exd C_{1}
 \quad \hbox{and} \quad
 F_{4} = \exd C_{3} \,.
\ee

The scaling symmetry in this frame has the form $S^{\rm IIA} \to s^2\, S^{\rm IIA}$ if $e^{-\phi} \to s \, e^{-\phi}$, $C_\ssM \to s \, C_\ssM$ and $C_{\ssM\ssN\ssP} \to s \, C_{\ssM \ssN \ssP}$, with $\hat g_{\ssM \ssN}$ and $B_{\ssM \ssN}$ held fixed. So once again we expect the on-shell action to evaluate to a boundary term, and ask what this boundary term is. We identify the boundary term in the Einstein frame, obtained by the Weyl scaling $\hat{g}_{\ssM \ssN} = e^{\phi/2} g_{\ssM \ssN}$, since the field equations are simpler.

The Einstein-frame action becomes
\ba
 S^{\rm IIA} &=& - \frac{1}{2 \kappa_{10}^{2}}
 \int \exd^{10} x \sqrt{- g_{10}} \;
 \left[  \cR + \frac{1}{2} \, (\partial\phi)^{2}
 + \frac{e^{-\phi}}{2(3!)} \, H_{3}^{2}
 + \frac{e^{3\phi/2}}{2(2!)} \, F_{2}^{2}
 + \frac{e^{\phi/2}}{2(4!)} \, \tilde{F}_{4}^{2} \right] +S_{CS} \nn\\
 &=& - \frac{1}{2 \kappa_{10}^{2}} \int \left[ *  \cR
 - \frac{1}{2} \, \exd \phi \wedge *\exd \phi
 - \frac{e^{-\phi}}{2} \, H_{3} \wedge*H_{3}
 + \frac{e^{3\phi/2}}{2} \, F_{2} \wedge *F_{2} \right.\\
 && \qquad\qquad\qquad\qquad\qquad\qquad\qquad\qquad
 \left. + \frac{e^{\phi/2}}{2} \,
 \tilde{F}_{4} \wedge *\tilde{F}_{4}
 + \frac{1}{2} \, B_{2} \wedge F_{4} \wedge F_{4}
 \right] \,, \nn
\ea
leading to the following equations of motion for the form fields
\ba
 \exd \Bigl( e^{-\phi} *H_{3} + e^{\phi/2} \, C_{1} \wedge *\tilde{F}_{4} \Bigr) &=& - \frac{1}{2} \,
 F_{4} \wedge F_{4} \\
 \exd \Bigl( e^{3\phi/2} \, *F_{2} \Bigr)
 &=& - e^{\phi/2} \, H_{3} \wedge *\tilde{F}_{4} \\
 \exd \Bigl( e^{\phi/2} \, *\tilde{F}_{4}
 +F_{4} \wedge B_{2} \Bigr) &=& 0 \,.
\ea
The trace of the Einstein equations similarly gives
\be \label{IIAEtrace}
 - \cR = \frac{1}{2} \, (\partial\phi)^{2}
 + \frac{e^{-\phi}}{4(3!)} \, H_{3}^{2}
 + \frac{3e^{3\phi/2}}{8(2!)} \, F_{2}^{2}
 + \frac{e^{\phi/2}}{8(4!)} \, \tilde{F}_{4}^{2} \,.
\ee

Substituting eq.~\pref{IIAEtrace} into the action eliminates the curvature scalar,
\begin{equation}
 S^{\rm IIA}_{\rm on-shell} = - \frac{1}{4 \kappa_{10}^{2}}
 \int \left( - \frac{e^{-\phi}}{2} \, H_{3} \wedge *H_{3}
 + \frac{e^{3\phi/2}}{4} \, F_{2} \wedge *F_{2}
 + \frac{3e^{\phi/2}}{4} \, \tilde{F}_{4}
 \wedge *\tilde{F}_{4}
 + B_{2} \wedge F_{4} \wedge F_{4} \right) \,,
\end{equation}
which can be written as a total derivative using the form-field equations of motion:
\ba
 S^{\rm IIA}_{\rm on-shell} &=& -\frac{1}{8 \kappa_{10}^{2}}
 \int \exd \left( - e^{-\phi} \, B_{2} \wedge *H_{3}
 + \frac{e^{3\phi/2}}{2} \, C_{1} \wedge *F_{2}
 + \frac{3e^{\phi/2}}{2} \, C_{3} \wedge *\tilde{F}_{4}
 \right. \\
 && \qquad\qquad\qquad\qquad\qquad\qquad
  \left. - e^{\phi/2} \, B_{2}
 \wedge C_{1} \wedge *\tilde{F}_{4}
 + \frac{3}{2} \, C_{3} \wedge F_{4} \wedge B_{2}
 \right) \,. \nn
\ea

\subsection*{10D Type IIB supergravity}
\label{tobe}

The starting point is the bosonic part of the Type IIB lagrangian density in 10D, which again involves the NS-NS fields $\phi$, $g_{\ssM\ssN}$ and $B_{\ssM \ssN}$; the Ramond-Ramond gauge potentials $C$, $C_{\ssM \ssN}$ and $C_{\ssM \ssN \ssP \ssQ}$.

The string-frame lagrangian for these fields is \cite{GKP}
\ba \label{JFlagr}
 \cL^{\rm IIB}
 &=& - \frac{1}{2\kappa_{10}^2} \, \sqrt{- \hat g}
 \left\{ e^{-2\phi} \, \hat g^{\ssM\ssN}
 \Bigl( \hat \cR_{\ssM \ssN} + 4 \, \partial_\ssM \phi \, \partial_\ssN \phi \Bigr)
 + \frac12 \, \hat g^{\ssM \ssN} F_\ssM F_\ssN \right. \nn\\
 && \qquad + \frac{1}{12} \, \hat g^{\ssM \ssN}
 \hat g^{\ssP \ssQ} \hat g^{\ssR \ssS}
 \Bigl[ \widetilde F_{\ssM \ssP \ssR}
 \widetilde F_{\ssN \ssQ \ssS}
 + e^{-2\phi} H_{\ssM \ssP \ssR} H_{\ssN \ssQ \ssS}
 \Bigr] \nn\\
 && \qquad\qquad + \left. \frac{1}{480} \, \hat g^{\ssM \ssN}
 \hat g^{\ssP \ssQ} \hat g^{\ssR \ssS}
 \hat g^{\ssT \ssU} \hat g^{\ssV \ssW}
 \widetilde{F}_{\ssM \ssP \ssR \ssT \ssV}
 \widetilde{F}_{\ssN \ssQ \ssS \ssU \ssW}
 \right\} \nn\\
 && \qquad\qquad \qquad - \frac{i}{8\kappa^2}
 \, e^\phi \, C_{(4)} \wedge \Bigl(
 \widetilde F_{(3)} \wedge \widetilde F_{(3)}
 + e^{-2\phi} H_{(3)} H_{(3)} \Bigr) \,,
\ea
where $F_{(k+1)} := \exd C_{(k)}$ is the field strength for the $k$-form Ramond-Ramond gauge potentials, $C_{(k)}$, and $H_{(3)} := \exd B_{(2)}$. The $\widetilde F$'s are defined by
\be \label{app:tildeFdefs}
 \widetilde F_{(3)} := F_{(3)} - C H_{(3)}
 \quad \hbox{and} \quad
 \widetilde F_{(5)} = * \widetilde F_{(5)} := F_{(5)} - \frac12 \,
 C_{(2)} \wedge H_{(3)} + \frac12 \, B_{(2)} \wedge F_{(3)} \,.
\ee

This lagrangian scales as $\cL^{\rm IIB} \to s^2 \, \cL^{\rm IIB}$ if the fields are scaled as follows
\be \label{app:scalingdefIIB}
 e^{-\phi} \to s \, e^{-\phi} \,, \quad
 \hat g_{\ssM \ssN} \to \hat g_{\ssM \ssN} \,, \quad
 C_{(k)} \to s C_{(k)} \,, \quad
 B_{(2)} \to B_{(2)} \,.
\ee
This is most easily seen from eq.~\pref{JFlagr} since each term but the last is quadratic either in $e^{-\phi}$ or one of the $\widetilde F$'s, and \pref{tildeFdefs} shows that $\widetilde F_{(k)} \to s \, \widetilde F_{(k)}$ under the transformation \pref{app:scalingdefIIB}. The last term is cubic in these fields but also has a compensating factor of $e^\phi$ out front. We are again guaranteed that the lagrangian becomes a total derivative on shell.

To identify what the derivative is, it is more convenient to use the Einstein frame, $\hat g_{\ssM \ssN} := e^{\phi/2} g_{\ssM \ssN}$, and to group the other fields into the complex quantities that transform simply under $SL(2,R)$,
\be \label{tauGdefs}
 \tau := C + i \, e^{-\phi}
 \quad \hbox{and} \quad
 G_{(3)} := F_{(3)} - \tau \, H_{(3)} \,.
\ee
In terms of these the lagrangian density becomes
\bea \label{app:EFlagr}
 \cL &=& - \frac{1}{2\kappa_{10}^2} \sqrt{-g} \; \left[ g^{\ssM \ssN} \left(
 \cR_{\ssM \ssN} + \frac{\partial_\ssM \tau \partial_\ssN \overline \tau}{2
  \left(\hbox{Im} \, \tau \right)^2} \right)
  + \frac{1}{12 \, \hbox{Im} \, \tau} \, G_{\ssM \ssP \ssR} \overline{G}^{\ssM \ssP \ssR} \right. \nn\\
 && \qquad \left.
 + \frac{1}{480} \, \widetilde{F}_{\ssM \ssP \ssR \ssT \ssV}
 \widetilde{F}^{\ssM \ssP \ssR \ssT \ssV} \right]
  - \frac{i}{8\kappa_{10}^2} \frac{ C_{(4)} \wedge
 G_{(3)} \wedge \overline G_{(3)}}{\hbox{Im} \, \tau}  \,,
\eea
and the scaling of the complex fields becomes
\be
 \tau \to s \, \tau \quad \hbox{and} \quad
 G_{(3)} \to s \, G_{(3)} \,.
\ee

To identify the on-shell action eliminate the Ricci scalar using the trace of the Einstein equations\footnote{ $F_{5}^{2}$ vanishes because the five-form is self-dual.}
\be
  -\cR =  \frac{\partial_\ssM \tau \partial^{\ssM} \overline \tau}{2
  \left(\hbox{Im} \, \tau \right)^2} +
  \frac{ G_{\ssM \ssN \ssP}
  \overline{G}^{\ssM \ssN \ssP}}{24 \,
  \hbox{Im} \, \tau  } \,.
\ee
Used in the action this yields
\bea
 S^{\rm IIB}_{\rm{on-shell}} &=&
 - {1 \over 48 \kappa_{10}^{2}}
 \int d^{10}x  \sqrt{-g_{10}} \; \frac{ {G_{3} \cdot
  \overline{G}_{3}}}{\hbox{Im} \, \tau  }
  - {1 \over 4 \kappa^{2}_{10}}
  \int C_4 \wedge H_3 \wedge F_3 \nn\\
   &=& - {1 \over 2 \kappa^{2}_{10}} \int   \bigg( {1 \over 4} e^{\phi} \widetilde{F}_{3} \wedge  * \widetilde{F}_{3}
     +{1 \over 4} e^{-\phi} {H}_{3} \wedge  * {H}_{3} \bigg) - {1 \over 4 \kappa_{10}^2} \int C_4
     \wedge H_3 \wedge F_3 \\
    &=& -{1 \over 2 \kappa^{2}_{10}} \int \bigg[ {1 \over 4} e^{\phi} {F}_{3} \wedge  * \widetilde{F}_{3}
     +{1 \over 4}  {H}_{3} \wedge  \Bigl( e^{-\phi}*{H}_{3}  - C_{0} e^{\phi}  * \widetilde{F}_{3}
     \Bigr)  + {1 \over 2} \, C_4 \wedge H_3
     \wedge F_3 \bigg]\,.\nn
\eea
Integrating by parts gives
\ba
 S^{\rm IIB}_{\rm{on-shell}} &=& -
 {1 \over 2 \kappa^{2}_{10}} \int  \bigg\{ {1 \over 4} \,
 \exd \bigg(  {C}_{2} \wedge   e^{\phi}
 *\widetilde {F}_{3} \bigg)
 + {1 \over 4}  \, \exd \bigg[ {B}_{2} \wedge \Bigl( e^{-\phi}*{H}_{3} - C_{0} e^{\phi} * \widetilde{F}_{3} \Bigr) \bigg]  \cr
 && \qquad\qquad\qquad -  {1 \over 4} \, {C}_{2} \wedge \exd
 \bigg(   e^{\phi}*\widetilde {F}_{3} \bigg)
 - {1 \over 4} \, {B}_{2} \wedge \exd
 \bigg[ \Bigl( e^{-\phi}*{H}_{3}  - C_{0} e^{\phi}  * \widetilde{F}_{3} \Bigr) \bigg] \cr
 && \qquad\qquad\qquad\qquad + {1 \over 2} \, C_4 \wedge H_3 \wedge F_3 \bigg\} \,.
\ea
Next we use the three-form field equations,
\ba
 \exd \bigg(  e^{\phi}* \widetilde{F}_{3} \bigg) &=&  \widetilde{F}_{5} \wedge H_3 \cr
 \exd \bigg( e^{-\phi}*{H}_{3}  - C_{0} e^{\phi}  * \widetilde{F}_{3} \bigg) &=&  F_{3} \wedge
 \widetilde{F}_{5} \,,
\ea
to write
\ba
 S^{\rm IIB}_{\rm{on-shell}} &=& -
 {1 \over 2 \kappa^{2}_{10}} \int \bigg\{
 {1 \over 4}\, \exd \bigg(  {C}_{2} \wedge
  e^{\phi} *\widetilde {F}_{3} \bigg)
   +{1 \over 4} \, \exd \bigg[ {B}_{2} \wedge  \Bigl( e^{-\phi}*{H}_{3}  - C_{0} e^{\phi}
   * \widetilde{F}_{3} \Bigr) \bigg] \nn\\
 && \qquad \qquad - {1 \over 4} \, {C}_{2}
 \wedge \widetilde{F}_{5} \wedge H_3
 - {1 \over 4} \, {B}_{2} \wedge F_3 \wedge
 \widetilde{F}_{5}
  + {1 \over 2} \, C_4 \wedge H_3
  \wedge F_3 \bigg\} \nn \\
 &=& - {1 \over 8 \kappa^{2}_{10}}
 \int \exd \bigg[    {C}_{2} \wedge   e^{\phi}
 *\widetilde {F}_{3}
 +  {B}_{2}
 \wedge  \Bigl( e^{-\phi}*{H}_{3}  - C_{0} e^{\phi}  * \widetilde{F}_{3} \Bigr)  \nn\\
 && \qquad\qquad\qquad \qquad\qquad \qquad  +
  C_4 \wedge C_2 \wedge H_3 -
   C_4 \wedge F_3 \wedge B_2 \bigg] \,.
\ea

\section{Solutions to the 10D metric/axio-dilaton equations}
\label{app:ftheorysolns}

We next briefly describe a situation where solutions are known fairly explicitly to the equations governing the metric and axio-dilaton in Type IIB supergravity. These are the unwarped, flat solutions of ref.~\cite{scs}.

\subsubsection*{Flat solutions}

When $n =2$ a very broad class of explicit solutions to the Einstein equations are known \cite{scs} in the limiting case where the two transverse dimensions are not warped: $\partial_m W =0$. In this case the $(\mu\nu)$ Einstein equation implies $R = 0$ and so the solutions are given by $\tau = \tau(z)$ and
\be
 \exd s^2 = \eta_{\mu\nu} \, \exd x^\mu \exd x^\nu
 + e^{2C(z,\overline z)} \, \exd \overline z \, \exd z \,.
\ee

A broad class of solutions to eq.~\pref{axiocomplex} are immediate when $\partial_m W = 0$ \cite{scs}: it is satisfied by any holomorphic function, $\tau = \tau(z)$, for which $\bar \partial \tau = 0$. The transformation properties of the axio-dilaton under the $PSL(2,Z)$ subgroup of the $PSL(2,R)$ symmetry are most easily tracked if $\tau(z)$ is written
\be
  j(\tau(z)) = P(z) \,,
\ee
where $j(\tau)$, is the standard bijection from the $PSL(2,Z)$ fundamental domain, $\cF$, to the complex sphere, given in terms of Eisenstein modular forms, $E_k(\tau)$, \cite{koblitz}. $P(z)$ is a holomorphic function whose singularities are chosen by the properties of the source branes.

The singularities of the metric turn out to be just conical at positions, $z = z_i$, where $P(z)$ has isolated poles. The metric turns out to be compact when $P(z)$ is a ratio of polynomials of equal degree whose numerator has 24 zeroes, such as for the choice
\be
 P(z) = \frac{4 (24 f)^3}{27 g^2 + 4 f^3} \,,
\ee
with $f(z)$ a polynomial of degree 8 and $g(z)$ a polynomial of degree 12. This gives a compactification of Type IIB supergravity on $CP^1$, corresponding to an F-theory reduction on $K3$ \cite{ftheory}.

The metric function $C(z,\overline z)$ is found by solving Einstein's equations, giving
\be \label{metricftheory}
 e^{2C(z,\overline z)} = (\hbox{Im}\, \tau) \, \left| \eta^2(\tau) \,
 \prod_{i = 1}^N (z - z_i)^{-1/12} \right|^2 \,,
\ee
where $\eta(\tau) = q^{1/24} \prod_k (1 - q^k)$, for $q = e^{2\pi i \tau}$, denotes the Dedekind $\eta$-function \cite{koblitz}, and the product runs over the singularities of $P(z)$.

Notice that because the $d$-dimensional metric is flat for all of these solutions, eq.~\pref{higherz} shows that any sources must satisfy $\hat g_{\mu\nu} (\delta S_{\rm source}/\delta \hat g_{\mu\nu})$ must vanish, at least when integrated over the Gaussian pillbox surrounding the source position. This turns out to be true, in particular, when $S_{\rm source}$ is the action of a $D7$-brane \cite{BBvN} or its image under $SL(2,Z)$.

Finally, the asymptotic form of $\tau(z)$ near the singularities may be found using the known properties of $j(\tau)$. In particular, for large $\hbox{Im}\, \tau$, $j(\tau) \simeq e^{-2\pi i \tau} + \cdots$ and so where $P(z) \simeq c_i/(z - z_i)$ the above solution implies
\ba
 \tau(z) &\simeq& \frac{1}{2\pi i} \, \ln (z-z_i) + \cdots \nn\\
 \hbox{and} \quad
 e^{2C(z,\overline z)} &\simeq& k \; \hbox{Im} \,
 \tau \simeq - \frac{k}{2\pi} \, \ln |z - z_i| + \cdots\,,
\ea
as $z \to z_i$, for $k$ a positive constant.

%%%%%%%%%%%%%%%%%%%%%%%%%%%%%%%%

%%%%%%%%%%%%%%%%%%%%%%%%%%%%%%%%%%%%
\end{document}